# Silicon carbide-assisted co-existence of magnetic phases in well-optimized Ti$_3$SiC$_2$-etched MXene

Qandeel Noor,[1] Syedah Afsheen Zahra,[1] Syed Rizwan[1]*

Physics Characterization and Simulations Lab (PCSL), Department of Physics, School of Natural Sciences (SNS), National University of Sciences and Technology (NUST), Islamabad 44000, Pakistan

Corresponding Author: Syed Rizwan: syedrizwan@sns.nust.edu.pk; syedrizwanh83@gmail.com; Tel: +92 90855599

## Abstract

Here, we report the first successful exfoliation of two-dimensional Ti$_3$C$_2$T$_x$ MXene through selective etching of silicon from titanium silicon carbide (Ti$_3$SiC$_2$) MAX. The successful etching and exfoliation of MXene is confirmed through the shifting of all (00l) peaks to lower angles along with the increase in c-lattice parameter as determined by X-ray diffraction technique to detail the material structure. The c-lattice parameter of multilayered MXene was found to be 19.34Å which was increased to 26.22 Å after delamination process indicating the successful intercalation of TMA$^+$ ions within the MXene Sheets. The scanning electron microscopy (SEM) images show the formation of 2D layered structure. The magnetic measurement of the etched MXene sample was measured using superconducting quantum interference device (SQUID: Quantum Design). The magnetization vs magnetic (M-H) curves clearly indicate the ferromagnetic-dominant hysteresis loops at low-temperature as well as at room-temperature along with the presence of small diamagnetic phase due to the presence of silicon carbide (SiC) present in MXene structure. The presence of SiC phase is confirmed through XRD and Raman spectra that show the sharp peaks and vibrational modes of SiC within 2D MXene structure. The present work shows the co-existence of ferromagnetic and diamagnetic phases making it suitable 2D material for future spintronics devices.

## 1. Introduction

Soon after the discovery of graphene, two-dimensional (2D) materials attracted much attention owing to their enormous potential [1-5]. The discovery of new 2D material namely MXene, that belongs to the family of Transition metal carbides with open d-orbital shell, could reveal exciting properties due to its diverse oxidation and spin-states, and greater spin-orbit coupling. The richness in elemental composition and chemical decoration offers an outstanding ground for developing and discovering internal degrees of freedom of electron-charge, spin and orbital motion and their relationship for fundamental exploration in advanced spintronics application [6].

MXene is derived from three-dimensional (3D) MAX phases using elegant exfoliation approach. The general formula for MXene is $M_{n+1}X_nT_x$, where, M stands for early transition metal (Ti, Nb, Ta, Mo, V, etc), X can be either C and/or Ni and $T_x$ represents the functional groups attached after chemical exfoliation of MAX is performed. The general formula for MAX phase is $M_{n+1}A_nX_n$ where, A represents the elements of group IIIA and IVA (Si, Al , Ge, Sn, S, P, As, etc) and subscript 'n' is an integer (n=1,2,3) with 211, 312, and 413 MAX Phases [1-6]. High surface area nano-porous MXenes can be prepared by using wet chemical etching method which is convenient in terms of yield, feasibility, cost-effectiveness and controllability [7]. In wet chemical etching method, acidic fluoride containing aqueous solutions are extensively used for exfoliation of MXene phases from their respective MAX precursors. More than 80 different types of MAX phases are still under process and thus far, more than 30 different MXenes are procured experimentally and many more are expected to be produced [6,7]. MXenes are equipped with highly variable and alterable chemical and physical properties because of the richness of stoichiometry and surface functional groups. For instance, high thermal stability, large electrical conductivity, tunable bandgap, large elastic moduli, capability to intercalate ions and hydrophilicity makes MXene the most promising 2D material for future applications [8].

The most substantially studied MXene is titanium Carbide-$Ti_3C_2$ which is the first MXene derived from $Ti_3AlC_2$ in 2011. Most of the MXenes reported until now are derived only from Al containing MAX phases. This is because the intrinsic reduction potential of Al is less compared to the other elements of the same group [8,9]. The $Ti_3AlC_2$ is the only MAX precursor used by the researchers in early years but due to its high manufacturing cost and limited availability, it is suggested to be replaced by other available MAX phases such as $Ti_3SiC_2$. Interestingly, due to different magnetic nature of Si compared to Al, the resulting magnetic behavior of MXene can be different [9].

In our study, we have considered $Ti_3SiC_2$ MAX phase as the parent compound to produce the desired $Ti_3C_2$ MXene phase. In 2010, the first synthesis of $Ti_3C_2$ from $Ti_3SiC_2$ was experimented but the attempt failed because the bonding between Ti-Si in $Ti_3SiC_2$ is stronger compared to the bonding between Ti-Al in $Ti_3AlC_2$ and Ti-Ge in $Ti_3GeC_2$ [10]. Due to the strong bonds between Ti-Si, the $Ti_3SiC_2$ MAX shows remarkable resistance towards strong bases and acids including HF [12]. Thus, for the wet chemical etching of Si along with mixture Hydrofluoric acid (HF), some oxidizing agent is also required because oxidants can abet in breaking the very strong bonds of Si with titanium and with other elements such as Si-C bonds in a very selective way [11,12].

Here, we have reported the successful etching of Si from $Ti_3SiC_2$ using wet chemical etching method however, a small amount of Si is present in the sample in the form of SiC phase which is used to induce the diamagnetic phase in the MXene sample thus, resulting in co-existence of ferromagnetic and diamagnetic phases in our best optimized sample. The structural, morphological, optical, electronic and magnetic properties are extensively discussed for our best optimized multilayered MXene sample. This is the first report which shows the successful delamination of $Ti_3C_2$ MXene extracted from $Ti_3SiC_2$ MAX compound that shows co-existence of two magnetic phases contrary to the previous reports on the presence of ferromagnetic phase in $Ti_3C_2$ MXene etched from $Ti_3AlC_2$ MAX phase and give flexibility in the chemically-derived 2D MXenes.

## 2. Experimental Procedure

### 2.1 Materials

Commercially available $Ti_3SiC_2$ MAX powder (mesh size: 200 – 325, particle size <250um, purity > 98%) was used to produce $Ti_3C_2$ MXene. Hydrofluoric acid (HF, 30 wt. %, 17M) was used for chemical etching. Ultra-pure deionized water was used for washing purpose and hydrogen peroxide ($H_2O_2$, 35 wt. %, 11.7M) was used as an oxidizing agent during etching Process. Paraffin oil was used in an oil bath to provide uniform heating to the etchant mixture.

### 2.2 Optimized of chemical etching and preparation of $Ti_3C_2Tx$ MXene

The etching solution of MXene was prepared by adding 5 mL of $H_2O_2$ and 45 mL of HF in Teflon Beaker. The whole mixture of HF/$H_2O_2$ was placed on an ice bath for continuous stirring over the course of 30 minutes before adding 3g of $Ti_3SiC_2$ MAX powder. The temperature of the ice bath must be maintained at ≤ 5°C as the etching process is extremely exothermic and $H_2$ gas would release during the reaction hence, the lid of the bottle is kept loose in order to let the gas release during the process. The Teflon beaker carrying the solution was transferred to an oil bath pre-set at 30 °C for a continuous etching process for 46 hours. After this, the reaction was stopped. The washing of MXene was done in 5 cycles using 45 mL of centrifuge tubes. After each cycle, water-like supernatant was drained out and 50 mL of fresh deionized water was added to continue washing further. After the 5$^{th}$ cycle, $Ti_3C_2$ MXene sludge was washed with 50 mL of deionized water by using vacuum-assisted filtration with additional 100 mL of deionized water. After the successful washing process, $Ti_3C_2$ MXene flakes were collected and dried in vacuum oven at 80 °C for 24 h.

### 2.3 Delamination Processing of $Ti_3C_2T_X$ MXene

Without the addition of intercalant, MXene was delaminated by using probe-sonication technique. For this purpose, 1g of $Ti_3C_2$ MXene powder was disseminated in 50 mL of DI water followed by sonication under continuous cooling in ice bath. The sonication time was set for 60 min, the amplitude of probe sonicator was set to 40 % and probe sonication pulse time was set to 4 sec ON and 2 sec OFF. After the sonication, the colloidal solution of $Ti_3C_2$ MXene was centrifuged for 1 hour at 3500 rpm. Using this approach, higher concentration (1.3 mg/mL) of colloidal solution of $Ti_3C_2$ MXene was prepared. The flakes size of $Ti_3C_2$ MXene was also increased after probe sonication with no mass loss observed. In the second approach of delamination process, 1g of $Ti_3C_2$ powder was treated with 2 mL of tetra methyl-ammonium hydroxide (TMAOH) and by manual shaking for 03 mins; the wet powder of MXene turned into darker black color. In this dark black mixture of $Ti_3C_2$ MXene, 20 mL of DI water was added and allowed to stir for 24 hrs at 300 rpm. Washing of $Ti_3C_2$ MXene was done with DI water three times via centrifugation. The washing cycles are as follows: 10 mins for first two cycles and 15mins for last cycle at 3500 rpm. The dilute green solution of $Ti_3C_2$ MXene can be seen as a resilient after the first washing cycle. The PH of the basic $Ti_3C_2$ solution turned into neutral after the 3$^{rd}$ washing cycle. Freestanding MXene films can be synthesized by further washing of MXene with vacuum-assisted filtration. MXenes collected from both processes described above were dried under vacuum for 15h at room-temperature before further drying in vacuum oven at 120 °C.

## 3. Results and Discussion

### 3.1 Structural Analysis

X-ray diffractometer was used to study the structure of MXene in **Figure 1** which reveals the successful removal of silicon from $Ti_3SiC_2$ MAX phase after chemical treatment. After the removal of Si atoms from $Ti_3SiC_2$ MAX, the peak at 39º completely vanishes that shows the successful removal of Si layers from $Ti_3SiC_2$ resulting in the successful synthesis of $Ti_3C_2T_x$ MXene. Also, all the other peaks of $Ti_3SiC_2$ are shifted to lower angles that indicates the formation of $Ti_3C_2$ MXene and increase in the c-lattice parameter. The peak at 8.9º in the XRD of MXene is the major peak of $Ti_3C_2$ that is further shifted to lower angle after delamination Process [13-15]. Signatures of small phases of $Ti_3SiC_2$ are also observed indicating that a little quantity of Si in the form of SiC is still present after the etching process. The calculated c-lattice parameter of pure MAX is 17.63Å and after etching process, the c-lattice parameter increases to ~19.34Å. Similarly, delamination of $Ti_3C_2$ with Tetramethyl ammonium Hydroxide (TMAOH) shows the intercalation of TMA ions between the MXene [12]. The shift of (002) peak of $Ti_3C_2$ from 8.9º to 5.9º along with increase in the d-spacing from 9.87Å to 15.22Å confirms the delamination of $Ti_3C_2$ with TMAOH. The peak observed at 25º confirms the formation of $TiO_2$ that shows the little oxidation of MXene sample [13]. Few peaks at 34º and 36º show the SiC phase that is still present after successful exfoliation process. In the XRD patterns of $Ti_3SiC_2$, free-standing MXene films produced via vacuum-assisted filtration of TMA-insinuated and delaminated $Ti_3C_2$ colloidal solutions show crystalline and intense (00l) peaks as shown in **Figure 1**.

### 3.2 Surface Morphology

The surface morphology, studied using scanning electron microscopy (SEM) technique, show expanded layered structure after the silicon is removed from $Ti_3SiC_2$ that was also confirmed from increased c-lattice parameter [12,13] as shown in **Figure 2a-d**. SEM images indicate that the etching process is started from the edges of $Ti_3SiC_2$ that reveals the successful selective etching of Si from MAX phase. In the case of delaminated $Ti_3C_2$ MXene, more expansion between the MXene layers was observed due to the intercalation of TMA ions that indicates the accordion-like morphology of MXene layers [12,14].

Energy dispersive X-ray spectroscopy (EDX) allocates the information about elemental composition within 2D MXene nano-sheets after exposing it with $HF/H_2O_2$ as shown in **Figure 2e**

**& 2f,** respectively. The EDS results clearly indicates the presence of elemental peaks of Ti, Si,C,O,F and Al in both multi-layered MXene and MXene Films. The presence of Si and C in EDX confirms its SiC phase along with that Si is still present after achieving $Ti_3C_2$ MXene phase.

### 3.3 Raman Spectroscopy

Raman spectra of Multilayered $Ti_3C_2T_x$ MXene is shown in **Figure 3**. The existence of functional groups on MXene surface indicated by the presence of Raman-active phonon modes at different vibrational frequencies. As discussed by Didier[14] et al, Raman spectra of hexagonal unit cell of $Ti_3SiC_2$ (space goup p63/mmc) having seven vibrational modes out of total 33 optical modes, i.e, $2A_{1g}+2E_{1g}+3E_{2g}$ at different Phonon frequencies. The crystal spectrum shows well-defined phonon peaks at 159 $cm^{-1}$, 226 $cm^{-1}$, 279 $cm^{-1}$, 625 $cm^{-1}$, 673 $cm^{-1}$, along with two weak vibrational modes at 260 $cm^{-1}$ and 301 $cm^{-1}$. According to theoretical predication, the vibrational modes at 159 $cm^{-1}$ and 226 $cm^{-1}$ shows shear and longitudinal vibrations of Ti and Si atoms [14]. The phonon peak observed at 279 $cm^{-1}$ includes the shear vibrations of Si atoms with out-of-plane shear vibrations of adjacent Ti atomic planes. The phonon mode at 301 $cm^{-1}$ shows longitudinal vibrations of Si atomic planes with out-of-plane vibrations of adjacent Ti atoms. The vibrational modes at phonon frequencies at 625 $cm^{-1}$ and 673 $cm^{-1}$ that corresponds to $E_{2g}$ and $A_g$ modes, are the transverse and longitudinal modes involving adjacent C-atomic planes [14,18].

The unit cell of $Ti_3C_2T_x$ belongs to the point group $D_{3d}$ and phonon vibrations in the complete unit cell are described as: $4E_g+2A_{1g}+4E_u+2A_{2u}$. The phonon modes Eu and A2u are IR active modes while vibrational modes of Eg and A1g correspond to the Raman active modes. In the multilayered MXene (ML), 4Eg and 4Eu are doubly degenerated modes. This shows that there are four Raman active modes (Eg) that correspond to the in-plane vibrations of Ti and C atoms and $A_{1g}$, an out-of-plane vibrations of Ti and C atomic planes [15]. While Eu and A2u are in-plane and out-of-plane vibrational modes of Carbon planes before the attachment of Carbon atoms. After the removal of Si layers from $Ti_3SiC_2$, $Ti_3C_2T_x$ MXene surfaces can be terminated with functional groups (-O, -OH, -F). When terminal groups such as –O, -OH are attached onto the MXene surfaces, number of atoms in the unit cell increases that effects the lattice vibrations of MXene. Due to the cell distortion and attachment of surface terminal groups, phonon peaks are shifted and broadened [15,16].

The Raman spectrum and full peak assignment of multilayered MXene flakes, delaminated MXenes through TMAOH (TMAOH del) and MXene flakes through probe-sonication DI/PS del show multiple features in their peaks. In the Raman spectrum of multilayered $Ti_3C_2T_x$, A1g mode at 159cm$^{-1}$ of pure MAX phase that is compatible with vibrations of Ti-Si atoms, is shifted towards a low wavenumber (149cm$^{-1}$). Instead of Si, this vibration also involves atomic planes of C atoms and surface groups. The first peak associated with frequency 120cm$^{-1}$ is resonant peak which is coupled with plasmonic peak when 785nm laser light is used [17]. The next two phonon modes around 250cm$^{-1}$ and 400cm$^{-1}$ consisting of Eg (Ti,C,0) and $A_{1g}$ (Ti,C,O) are in-plane and out-of-plane vibrations of outer $Ti_2$ layer atoms as well as carbon atoms and surface groups, along with in-plane Eg vibrations of surface groups attached with Titanium atoms. The phonon peak observed at 660 cm$^{-1}$ confirms the Eg and $A_{1g}$ vibrations of Carbon atoms. The next two broad peaks between 1300 cm$^{-1}$ and 1650 cm$^{-1}$ correspond to the D-bands and G-bands; the D-band is associated with the formation of defects in MXene sample along with disordered graphite structure. Thus, the G-band arises due to the stretching of C-C bonds in Carbon based materials due to sp$^2$ sites [12-16]. Oxidation of $Ti_3C_2$ is also confirmed due to the presence of $TiO_2$ (at 149 cm$^{-1}$) particles on to the surface of disordered Carbon sheets that shows innermost Ti atoms migrated outward to react with Oxygen.

The Raman spectrum of delaminated MXene shows shifting in peak positions due to the intercalation of large molecules of TMAOH. When TMAOH is added after HF etching, $A_{1g}$ peak that corresponds to the carbon vibrations shifts to 700 cm$^{-1}$. The shifting in peak occur due to the increase in c-lattice parameter and inter layer spacing caused by large intercalant molecules [12,19]. The distortion in D and G -bands suggest that more defects are introduced between MXene sheets due to probe-sonication that dramatically exfoliates the delamination process and produced large fraction of high-quality 2D MXene.

3.4 X-ray Photoelectron Spectroscopy (XPS)

The actual surface chemistry of 2D multilayered $Ti_3C_2T_x$ MXene was analyzed by X-ray photoelectron spectroscopy (XPS). The XPS was carried out using a Kratos X-ray Photoelectron Spectrometer – Axis Ultra DLD with a silver anode beam to irradiate the sample surface. Figure 4(a-d) represents the deconvoluted XPS spectra of sputtered-clean $Ti_3C_2T_x$ MXene. The results were evaluated and analyzed on the basis of previous studies carried out by Halim et al. [20]. The bonding energies along with bonding types are presented in **Table SI**. **Figure 4a** represents the Ti2p region consisting of moieties I, II or IV and III mostly of Ti species Ti, Ti$^{+2}$ Ti$^{+3}$ while the moiety I refers to the bonding of M atoms with carbon atoms along with a birding oxygen, e.g. Ti$_3$C$_2$O$_x$ Moiety II denotes the bonding OH group to M bonded C atoms, e.g, Ti$_3$C(OH)$_x$. Moiety III signifies C and F bonding to M atoms, e.g, Ti$_3$C$_2$F$_x$. Moiety IV refers to the formation of OH-H$_2$O complexes with M atoms as a consequence of physiosorbed water molecules with M-OH termination, e.g, Ti$_3$C$_2$OH- H$_2$O, mentioned as H$_2$O$_{ads}$ Ti 2p (2p$_{1/2}$ and 2p$_{3/2}$) deconvoluted spectra consisting of energies for Ti (I, II or IV), TiO$_2$, TiO$_2$-F, C-Ti-F$_x$, C-Ti-T$_x$, C-Ti-O$_x$, C-Ti- (OH)$_x$, H$_2$O$_{ads}$ [21–27]. Peaks at binding energies (BE) of 454.6eV and 460.5eV represent the presence of Ti in Ti$_3$SiC$_2$ which arises due to the remanence of MAX phase as a result of incomplete etching or poor washing [28]. Figure 4b is a deconvoluted picture of Carbon 1s region fitted by four peaks. The largest peak is the one near to 282.1eV and approximately 78% is Ci-Ti-T$_x$. Peak at 285.3eV is attributed to SiC [29] and other two peaks are graphitic C-C or CH$_x$ or C-O bonds. The graphitic C-C arises due to the selective dissolution of Ti during the etching process [27]. The solvent used for washing and drying processes resulted in the formation of C-O and CH$_x$. O1s **Figure 4c** confirms the formation of oxides and hydroxides on the surface of the sputtered cleaned sheets. Peak appeared near 532.2eV is identified as SiO$_2$ corresponds to the residual silicon of synthesis material [32]. Presence of fluorine solely attributed to HF as shown in **Figure 4d** below.

### 3.5 Co-existence of Magnetic Phases

The electron dispersive spectroscopy (EDX) data as shown in **Figure 2b**. of Ti$_3$C$_2$T$_x$-MXene confirms the presence of different terminations (Tx= -O,-OH, -F) attached onto the MXene surface after the removal of Si layers from Ti$_3$SiC$_2$ MAX. As discussed by Sunaina et al [31], the reported EDS results shows the presence of little amount of M-atom, i.e. Aluminium (Si here) even after the completion of etching process. Aluminium itself is paramagnetic in nature [31] and its weak paramagnetic contribution does not affect much to the ferromagnetic nature of Ti$_3$C$_2$. In case of Ti$_3$SiC$_2$ MAX phase, successfully etched Ti$_3$C$_2$-MXene shows the presence of little amount of SiC phase as established by XRD, EDS and Rama spectra. The contribution of the diamagnetic phase of SiC can thus be expected in the etched Ti$_3$C$_2$T$_x$ MXene. The magnetization vs magnetic field (M-H) curves of Ti$_3$C$_2$ MXene taken at different temperatures are presented in **Figure 5a** which exhibits co-existence of ferromagnetic phase of Ti$_3$C$_2$ MXene accompanied by a small diamagnetic phase of SiC at all temperatures.

The M-H hysteresis loops for $Ti_3C_2$ MXene are measured at 5 K, 50 K, 300 K (**Figure 5a**). The $Ti_3C_2$ shows the ferromagnetic behavior along with slight introduction of diamagnetic phase at room-temperature (inset: Figure 5a). This small diamagnetic phase arises due to the presence of silicon Carbide phase. It is established that the $Ti_3C_2$ MXene is ferromagnetic in nature due to the delocalized electrons of Ti-3d states present around the Fermi level thus, the observed diamagnetic phase is related to the SiC phase. We suggest that the magnetic instability in the etched MXene arises due to the unpaired Ti-3d bonding states of external MXene sheet that causes increase in the electronic density of states near fermi level resulting in the redistribution of Ti-3d states from broken Ti-Si bonds into the delocalized Ti-Ti bonding placed near the Fermi level [32]. Shein et al [33] explained that the intra-layer MXene sheets are anti-ferromagnetic in nature while the magnetic behavior of MXene sheets occupying surface terminational groups such as ($T_x$= -O, -OH, -F) exhibits out-of-plane vibrations that makes it weak ferromagnetic at the edges. Thus, the Ti atoms on external MXene sheet have ferromagnetic ordering of spins [33,34]. The, the M-H loops of our etched sample indicates co-existence of conventional ferromagnetic phase and SiC-assisted diamagnetic phase in layered MXene the strength of which can be varied by varying the scale of selective removal of silicon atoms during the etching process. **Figure 5b** shows the magnetization vs. temperature (M-T) curve for $Ti_3C_2T_x$ MXene typical for a ferromagnet.

**Conclusion**

The 2D $Ti_3C_2$ –MXene sheets were extracted using wet-chemical etching method. The structural, morphological, optical, elemental, electronic and magnetic properties were studied using different characterization techniques that confirm the 2D nature of MXene. Raman spectra confirms the presence of vibrational modes associated with functional groups (-OH, -O, -F), along with the confirmation of D and G -bands present in carbon-based materials, and modes associated with SiC vibrations. The magnetic properties of $Ti_3C_2$ validates the co-existence of ferromagnetic and diamagnetic phases associated of $Ti_3C_2$ -MXene and SiC phases, respectively thus, making it possible for the simultaneous existence of two opposite magnetic phases. The presence of ferromagnetic/diamagnetic phases in two-dimensional layered MXene open up new doors for researchers to study two-dimensional materials for their possible application in spintronics.

**Acknowledgement**

The authors are thankful to Higher Education Commission (HEC) of Pakistan for providing research funding under the Project No. 6040/Federal/NRPU/R&D/HEC/2016. The author also thanks Office of Research and Innovation Center (ORIC) at National University of Sciences and Technology (NUST), Islamabad, Pakistan for their research support.

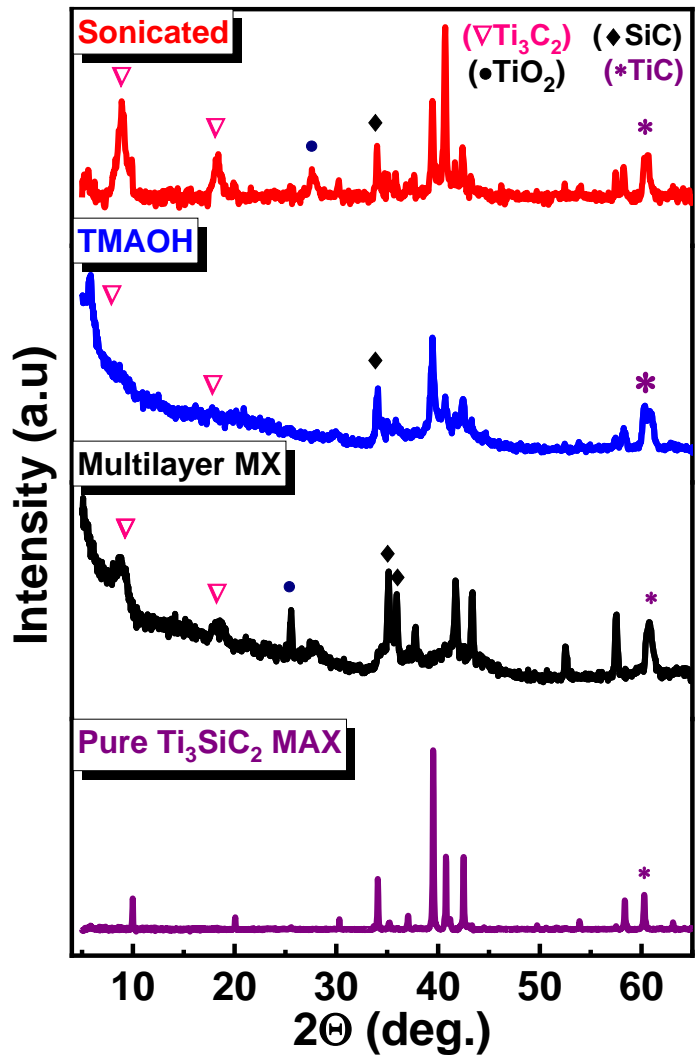

**Fig. 1.** The X-ray diffraction patterns of Ti₃SiC₂ MAX (purple), Multilayered Ti₃C₂ MXene (black), TMAOH-treated Ti₃C₂ MXene (blue) and Sonicated Ti₃C₂ MXene (red).

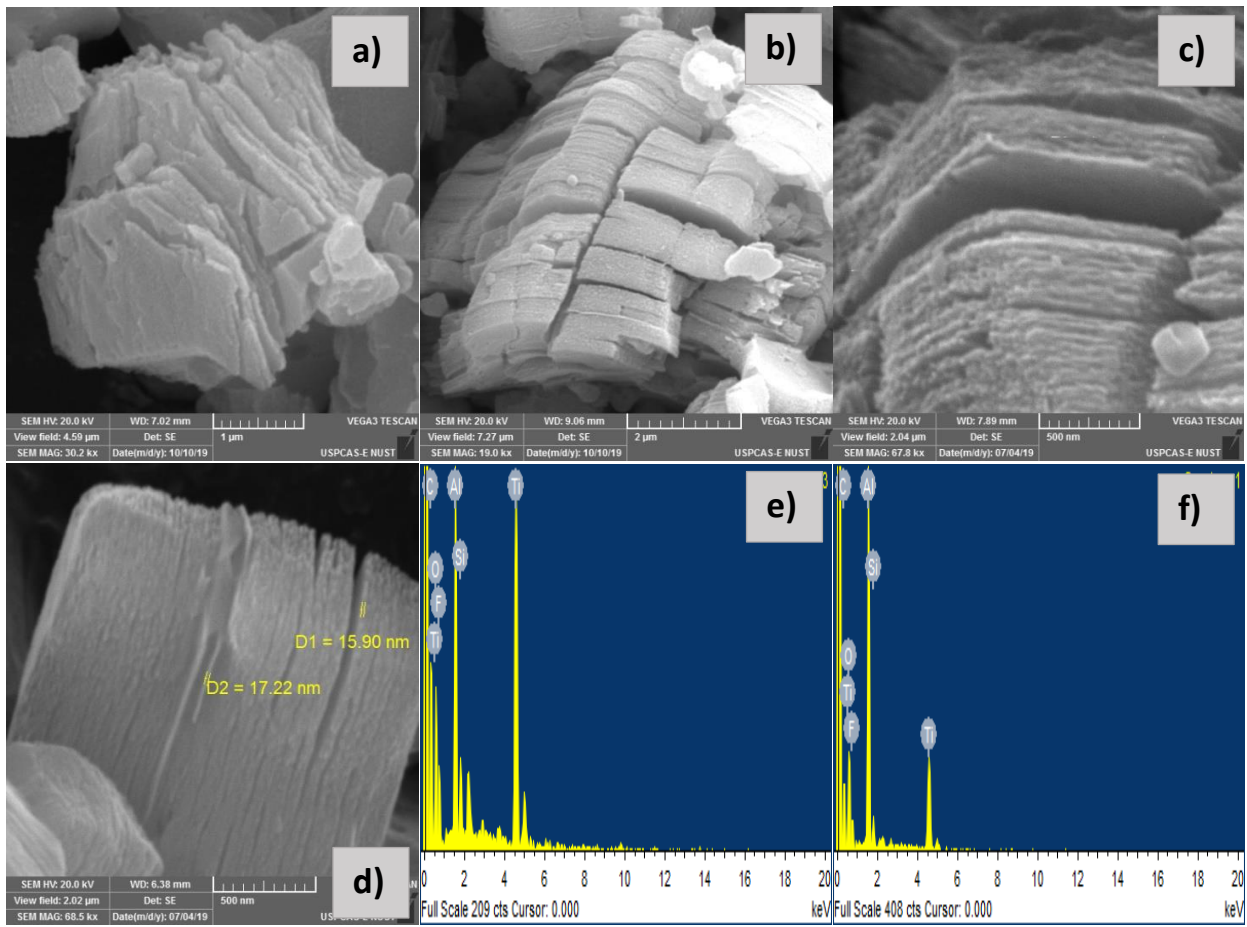

**Fig. 2. (a)**. Highly magnified SEM Images of $Ti_3SiC_2$ after treatment with $HF/H_2O_2$ exhibiting an expanding layers of $Ti_3C_2$ with open edges. **(b).** Energy dispersive X-ray spectroscopy (EDX) of MXene shows decrease in amount of Si after treating it with $HF/H_2O_2$.

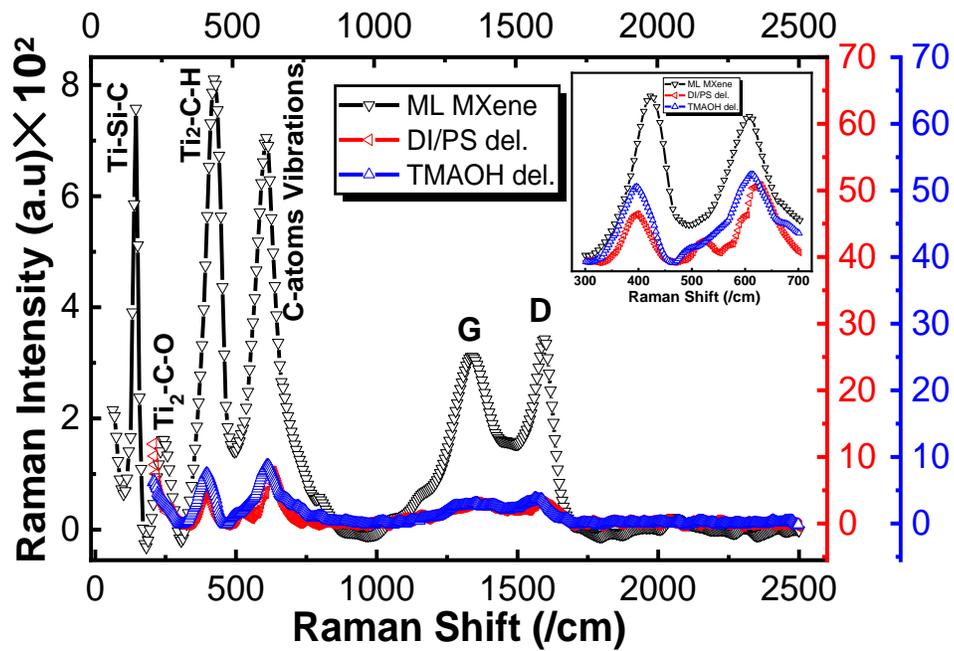

**Fig. 3.** Raman spectra of three samples at room-temperature.

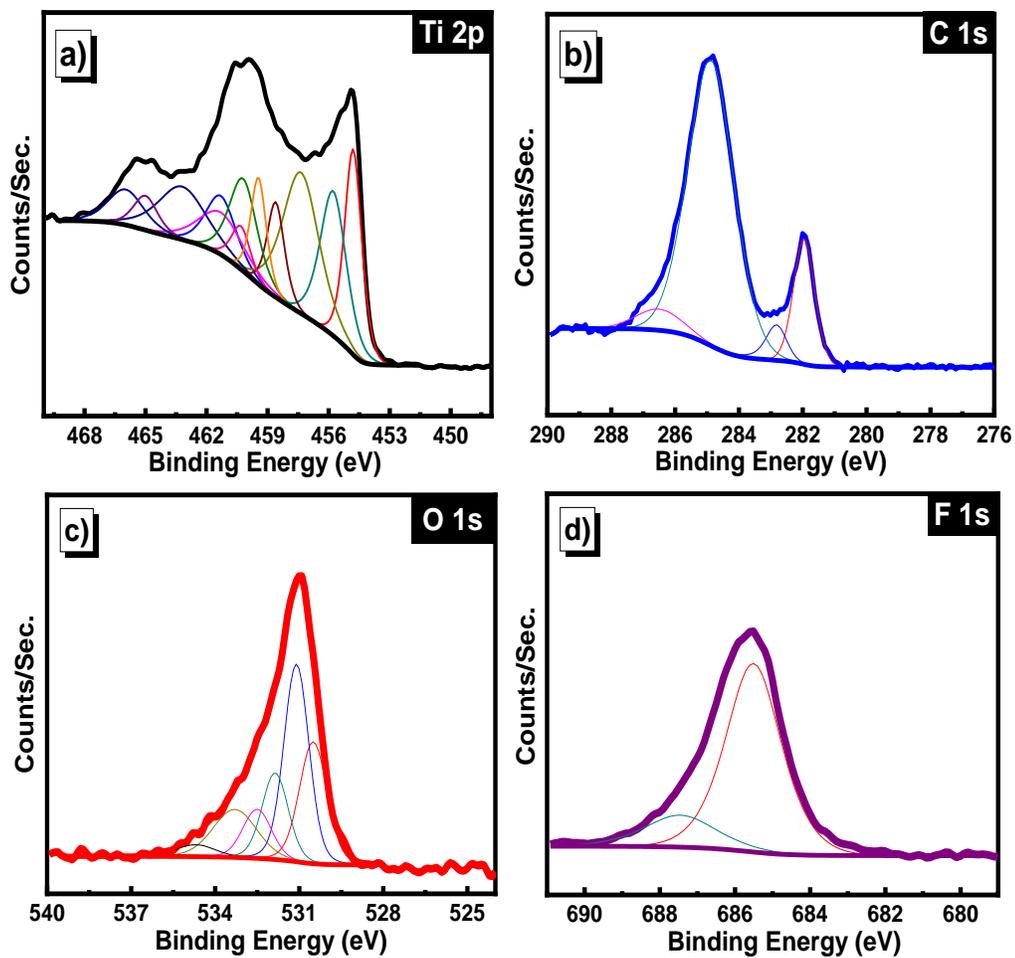

**Fig. 4.** XPS deconvoluted spectra of Multilayered $Ti_3C_2T_x$ MXene, **a)** Ti 2P **b)** C1s, **c)** O1s, **d)** F1s.

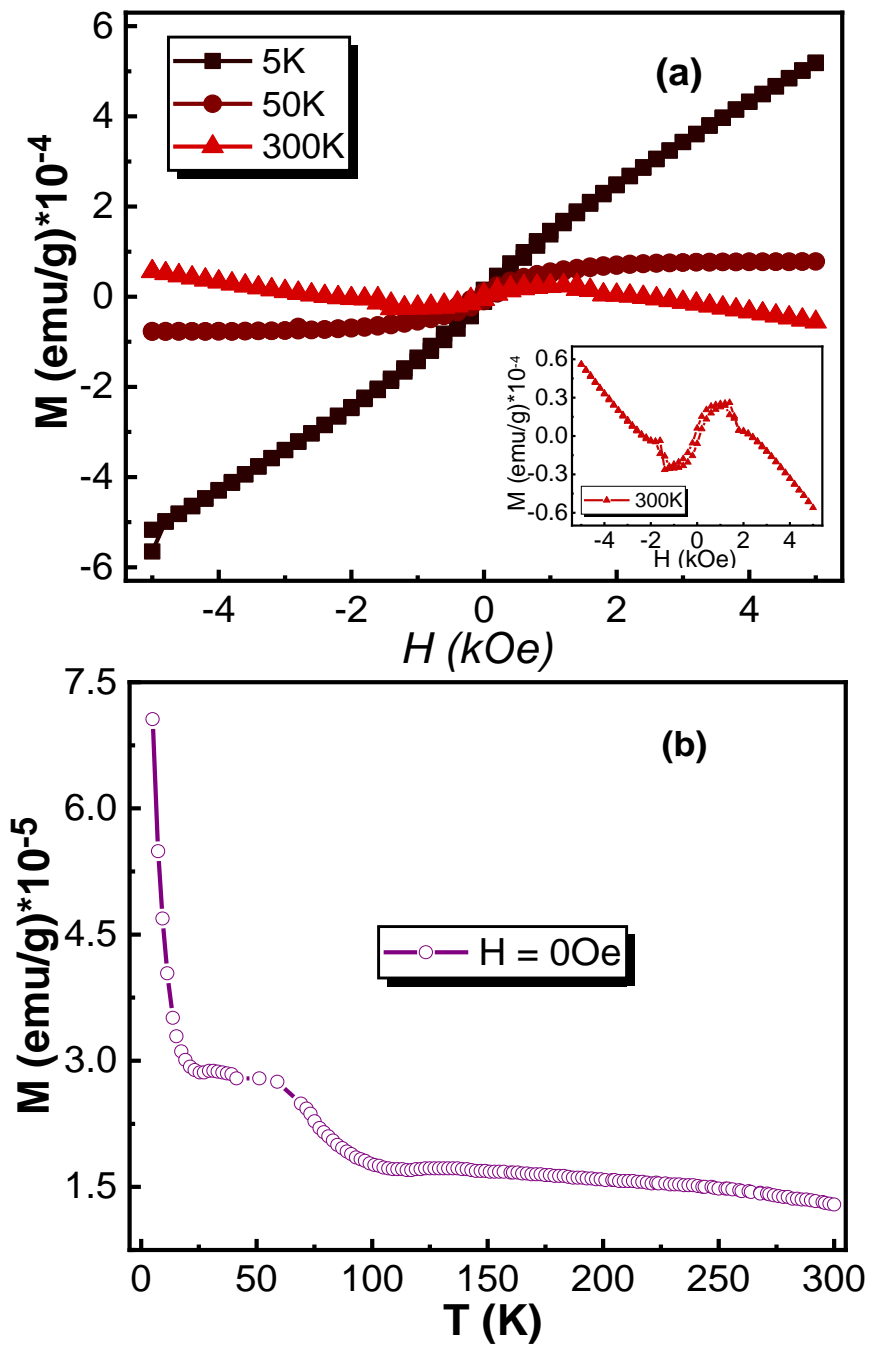

**Fig. 5. a)** The magnetization vs. magnetic field hysteresis measured at 5K, 50K, and 300K; the inset shows the magnified graph at 300K showing co-existence of diamagnetic and ferromagnetic phases. **b)** The magnetization vs temperature curves at zero applied magnetic field showing an increased magnetic moment at deceasing temperatures.